\documentclass[prb,preprint,showpacs]{revtex4}
\usepackage{graphicx}
\usepackage{amsmath,bm}
\usepackage{pstricks}
\usepackage{wasysym}

\begin{document}

\title{Twisted ultrathin silicon nanowires: a
possible torsion electromechanical nanodevice}

\author{Joelson Cott-Garcia and Joao F. Justo}

\affiliation{
  Escola Polit\'ecnica, Universidade de S\~ao Paulo,
CP 61548, CEP 05424-970, S\~ao Paulo, SP, Brazil
}

\begin{abstract}
Nanowires have been considered for a number of applications in nanometrology.
In such a context, we have explored the possibility of using ultrathin twisted
nanowires as torsion nanobalances to probe forces and torques at molecular
level with high precision, a nanoscale system analogous to the Coulomb's torsion
balance electrometer. In order to achieve this goal, we performed a first
principles investigation on the structural and electronic properties of twisted
silicon nanowires, in their pristine and hydrogenated forms. The results indicated
that wires with pentagonal and hexagonal cross sections are the thinnest stable
silicon nanostructures. Additionally, all wires followed a Hooke's law behavior
for small twisting deformations. Hydrogenation leads to spontaneous twisting,
but with angular spring constants considerably smaller than the ones for the
respective pristine forms. We observed considerable changes on the nanowire
electronic properties upon twisting, which allows to envision the possibility of
correlating the torsional angular deformation with the nanowire electronic
transport. This could ultimately allow a direct access to measurements on
interatomic forces at molecular level.
\end{abstract}

\maketitle

Low dimensional nanosystems yield large surface-to-volume ratios and quantum
confinement, properties that make them suitable to a number of potential
applications \cite{cui2001a,hayden2006,lieber2}, such as battery
electrodes, chemical
sensors, electron emission devices, and solar cells. Furthermore, nanostructured
materials could be used as tools in nanometrology \cite{metrol2012}, in order to
allow probing atomistic and molecular properties with unprecedented precision.
For instance, nanowires and nanotubes have already been used as probe tips
in scanning tunneling microscopy and atomic force microscopy to investigate
surface properties and manipulate matter at atomistic level.

Over the last decade, there has been great interest in growing one-dimensional
nanostructures with tailored mechanical, electronic, and optical
properties \cite{comini2009,growth}. Achieving such high degree of growth
control could allow, for example, obtaining nanowires for specific
applications in nanometrology. For this to become a reality in the coming years,
it is essential to acquire a thorough knowledge on the physical properties of
those nanostructures. Several wire properties could be explored in such
applications, such as electronic transport, thermal conductivity, optical
transitions, electron affinity, oxidation potential, and chemical reactivity.
It is also important to understand how nanowires deform and their elastic limit,
along with the correlation between mechanical and electronic properties. Several
other fundamental questions on nanowires emerge: what is the scaling law that
describes those properties? What is the smallest limit to get a stable ultrathin
nanowire? How do ultrathin nanowires behave under stretching and twisting
deformations? Here, we have addressed some of those questions by modeling the
properties of ultrathin twisted silicon nanowires and exploring their potential
applications in nanometrology as a torsion nanobalance \cite{cleland,kim}.

We performed a theoretical investigation on the structural and electronic
properties of ultrathin twisted silicon nanowires in their pristine and
hydrogenated forms, for several polygonal cross section shapes and
twisting angles. Our investigation focused on silicon nanowires,
although the conclusions could be extrapolated to wires of
several other chemical elements. Although silicon nanowires have been
widely investigated
\cite{cui2001b,ma,lauhon2004,hochbaum2008,rurali,amato2014},
there is scarce literature on the
properties of their twisted configurations \cite{sen,xu}. On the other hand,
twisted nanowires of several other materials have been grown and
studied recently \cite{popov,tizei,akaty}.

Our results indicated that nanowires with pentagonal and hexagonal cross
sections are the thinnest stable silicon wires in any twisted form.
Additionally, twisted nanowires, with any polygonal cross section, follow
the Hooke's law, for small stretching and twisting deformations around
their equilibrium configurations. In a pristine form, all wires were found
to be stable in an untwisted configuration. Upon hydrogenation, such
behavior changed substantially, with the respective equilibrium configurations
associated to highly twisted forms. All those results suggest that an
ultrathin nanowire could be used as a torsion nanobalance or electrometer
to probe forces and torques at molecular level \cite{naka}. The observed
changes on the nanowire electronic properties upon twisting could be
explored in terms of the electronic transport. This would lead to a direct
access to a measurement, in which the angular deformation could be
determined by the respective changes on the electronic current, allowing
to determine forces and torques involved in the interaction between nanowires
and molecular systems.

The calculations were performed using the Vienna {\it ab initio} simulation
package (VASP) \cite{vasp}. We considered nanowires with periodic boundary
conditions, with a tetragonal simulation cell. In the directions normal to
the nanowire, a lattice parameter of 20 \AA\ was used, which had enough
open space to prevent interactions between atoms in the original cell with
those in the neighboring images. In the nanowire direction ($z$), we
considered nanowires (for several polygonal cross sections and twisting
angles) with 10 layers, and the respective lattice parameter was
optimized according to an energy minimization iterative procedure. The
electronic exchange correlation potential was described within the density
functional theory/generalized gradient approximation \cite{pbe}. The spin
polarized electronic wave-functions were described by a projector augmented
wave method \cite{paw}, and expanded in a plane-wave basis set, with the
kinetic energy cutoff of 450 eV. For any system, convergence in total energy
was achieved when it differed by less than 10$^{-5}$ eV between two consecutive
self-consistent electronic calculation iterations. The structural relaxation,
for a nanowire in any configuration, was achieved when forces on the atoms
were smaller than $\rm 1\, meV/\AA$. All calculations took a
$1 \times 1 \times 8$ k-point mesh to sample the Brillouin zone.

\begin{figure}[tb]
\includegraphics[width=12cm,angle=270]{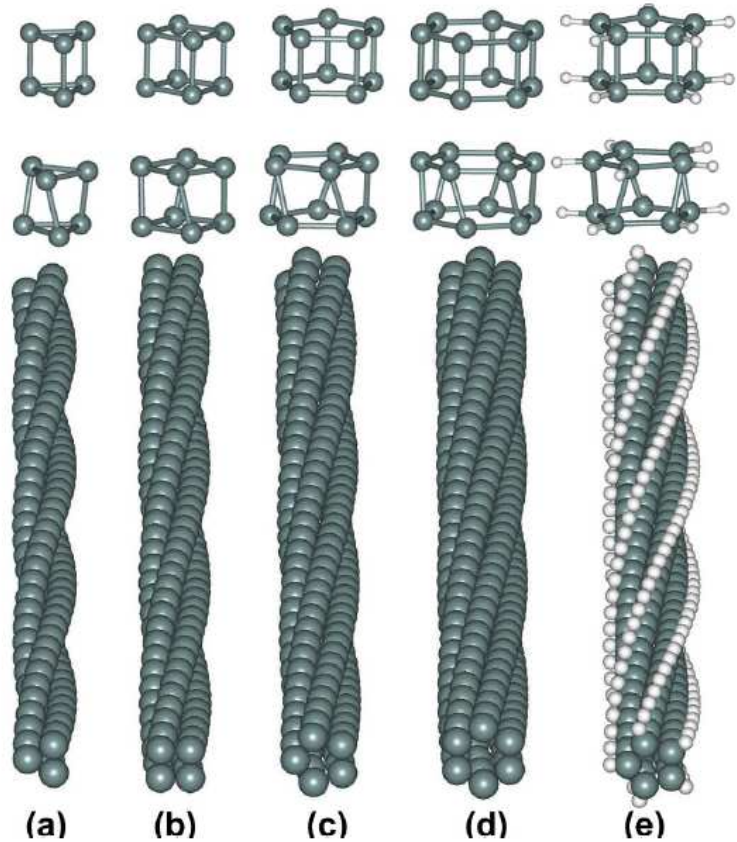}
\caption{Schematic representation of ultrathin silicon nanowires, in their
relaxed twisted configurations. The figure shows nanowires with (a) triangular,
(b) square,
(c) pentagonal, and (d) hexagonal cross sections in their pristine form.
The figure also shows (e) a hydrogenated pentagonal nanowire.
The top part of the figure presents the respective side views of two
neighboring planes. Gray and white balls represent silicon and hydrogen atoms,
respectively.}
\label{fig1}
\end{figure}

Figure \ref{fig1} presents a schematic representation of the ultrathin twisted
silicon nanowires studied here. We considered nanowires with several polygonal
cross sections, containing three, four, five, or six silicon atoms. Based on the
results of previous theoretical investigations, those structures are candidates
to be energetically stable \cite{li,sen,justo2007,justo2007b}.
Structures with larger cross
sections, i.e. containing more atoms in the rings, were found to be energetically
less favorable than those structures presented here. It should be stressed that
the systems investigated here can be classified as surface-like nanowire structures,
which differ from bulk-like ones \cite{ma,rurali}, in which the nanowires have a large core
(of at least 1 nm in diameter) with atoms in configurations equivalent to the respective
ones in the crystalline phase. In the last case, the nanowire properties
could be directly mapped to the respective properties in the crystalline phase. On the
other hand, for the ultrathin nanowires investigated here, such mapping is not that
direct.

\begin{figure}
\includegraphics[width=7cm]{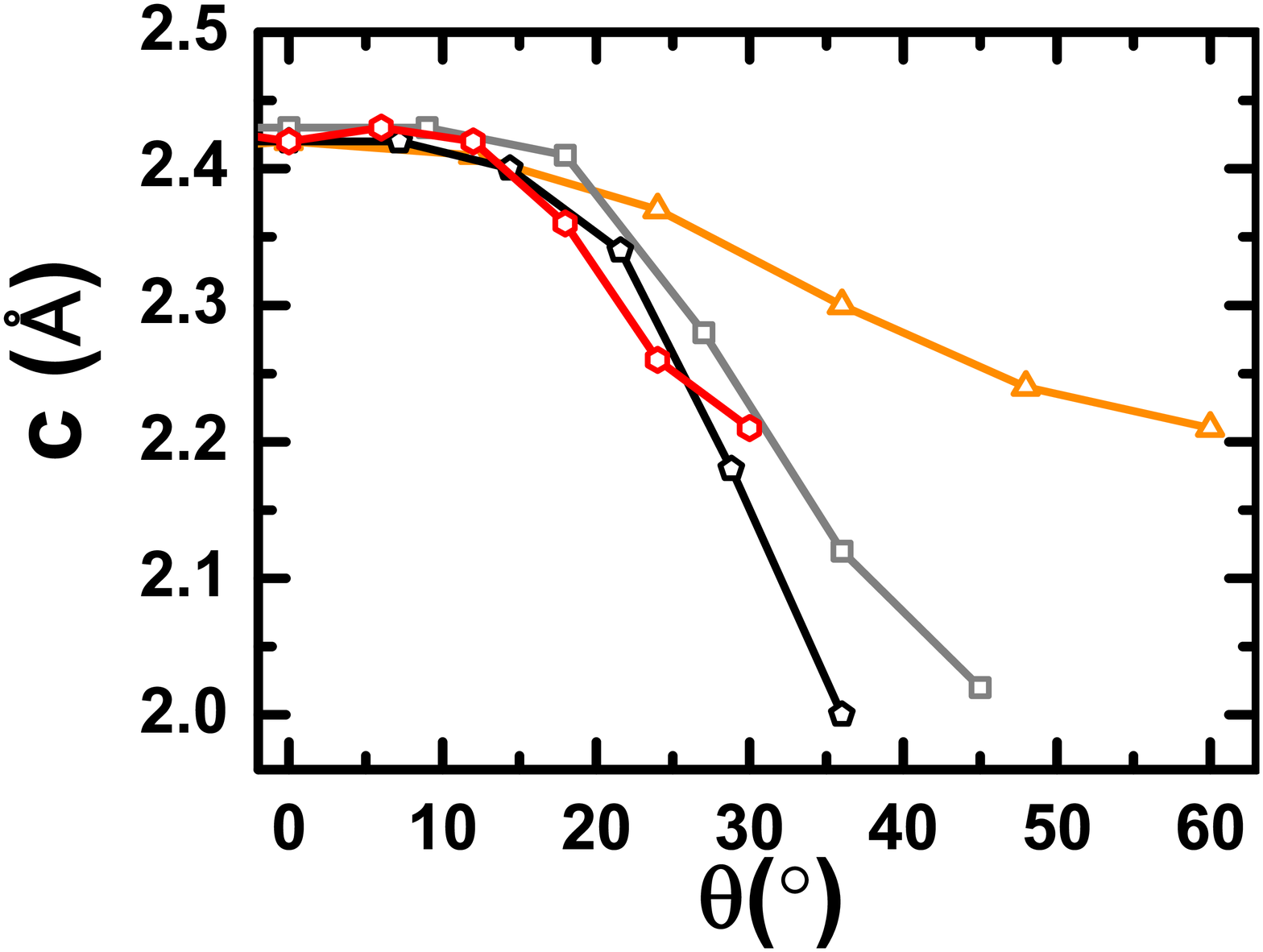}
\caption{Equilibrium parameter between two wire planes, $c$,
of nanowires (in pristine form) as a function of
the twisting angle $\theta$. The figure presents results for
nanowires with triangular ($\triangle$), square ($\Box$),
pentagonal ($\pentagon$), and hexagonal ($\hexagon$) cross sections.}
\label{fig2}
\end{figure}

Here, the twisting angle was
defined as the angle formed between two equivalent atoms in neighboring silicon
layers. Since we considered periodic boundary conditions in $z$-direction, there
was a limited number of possible intermediate twisting angles that could be
effectively studied, which was directly associated to the number of wire layers in
the simulation cell. Figure \ref{fig2} presents the distance (in the $z$ direction)
between two neighboring planes ($c$) as a function of the twisting angle for
pristine nanowires. The results indicated that, for all wire types, the largest
lattice parameter is associated to an untwisted configuration (0$^o$). In that case,
the lattice parameter is around 2.42 \AA\ for all wire types, which is a little
longer than the silicon interatomic distance in the crystalline diamond cubic
phase (of 2.37 \AA, computed within the same methodology and theoretical
approximations mentioned in the previous paragraphs). This could be explained
by the fact that, although silicon atoms in the nanowires are fourfold coordinated
as atoms in the crystalline phase, they have angles between bonds that
are far from the tetrahedral value, which causes important weakening on the
silicon-silicon bonds, as compared to the ones in the crystalline phase.

Twisting leads to a
substantial lattice parameter shortening, by as much as 20 \% (with respect to
untwisted configurations) for the largest respective twisting angles, which
corresponds, in the case of the pentagonal nanowire, to an angle of 36$^o$. This
lattice shortening, as a function of the twisting angle, comes simultaneously
with large increase in the in-plane interatomic distances, $\rm d(Si-Si)$, as
summarized in table \ref{tab1}. For example, in the pentagonal nanowire with
a twisting angle of 36$^o$, $\rm d(Si-Si)$  increases by about
11\% with respect to that in the untwisted configuration.

\begin{table}[ht]
\tiny
\caption{Properties of twisted silicon nanowires with several cross sections.
The table presents the properties as function of twisting angle ($\theta$):
interplanar distance (c), in-plane interatomic distance d(Si,Si),
Young's modulus ($\rm Y$), and {relative total energy per atom
($\rm \Delta E$/atom). Here,
total energies are given with respect to a reference value (-4.732 eV/atom for
pristine pentagonal and -4.020 eV/atom for hydrogen passivated pentagonal
silicon nanowires).}
Twisting angles are given in degrees ($^\circ$), distances in \AA~, energies in eV,
and Young's modulus in GPa.}
\label{tab1}
\begin{center}
\begin{tabular}{ccccccc}
  \hline
  shape &  $\theta$  &  c & d(Si-Si) &
 {$\rm \Delta E$/at}  & Y  \\
  \hline
{triangular} & 0  & 2.415 & 2.418 & {0.172} &   1316 \\
             & 12 & 2.409 & 2.420 & {0.176} &   1277 \\
             & 24 & 2.373 & 2.444 & {0.212} &   1153 \\
             & 36 & 2.298 & 2.469 & {0.207} &   855 \\
             & 48 & 2.238 & 2.477 & {0.201} &   417 \\
             & 60 & 2.209 & 2.473 & {0.205} &   1336 \\
  \hline
  {square} & 0  & 2.427 & 2.422 & {0.129} &   1300 \\
           & 9  & 2.434 & 2.428 & {0.131} &   1417 \\
           & 18 & 2.413 & 2.440 & {0.182} &   1081 \\
           & 27 & 2.277 & 2.508 & {0.200} &   1113 \\
           & 36 & 2.119 & 2.581 & {0.147} &   1136 \\
           & 45 & 2.024 & 2.641 & {0.085} &   1579 \\
  \hline
 {pentagonal}  & 0.0  & 2.421 & 2.405 & {0.00}  &   2412 \\
 {pristine}    & 7.2  & 2.415 & 2.397 & {0.030} &   2098 \\
               & 14.4 & 2.397 & 2.405 & {0.106} &   1751 \\
               & 21.6 & 2.337 & 2.424 & {0.147} &   1340 \\
               & 28.8 & 2.182 & 2.531 & {0.157} &   921 \\
               & 36.0 & 1.995 & 2.674 & {0.096} &   3422 \\
  \hline
{pentagonal} & 0.0  & 2.880 &  2.289 & {0.009} &  127 \\
{passivated} & 7.2  & 2.875 &  2.286 & {0.011} &  119 \\
             & 14.4 & 2.513 &  2.452 & {0.015} &  119 \\
             & 21.6 & 2.481 &  2.418 & {0.006} &  114 \\
             & 28.8 & 2.708 &  2.269 & {0.004} &  62\\
             & 36.0 & 2.651 &  2.258 & {0.00}  &  76 \\
  \hline
  {hexagonal} & 0  & 2.421 & 2.393 & {0.014} &  2809 \\
            & 6  & 2.433 & 2.390 & {0.035} &  2388 \\
            & 12 & 2.421 & 2.398 & {0.112} &  1931 \\
            & 18 & 2.355 & 2.419 & {0.143} &  1656 \\
            & 24 & 2.262 & 2.437 & {0.151} &  1195 \\
            & 30 & 2.205 & 2.474 & {0.150} &  1012 \\
\hline
\end{tabular}
\end{center}
\end{table}

\begin{figure}[tb]
\includegraphics[width=7cm]{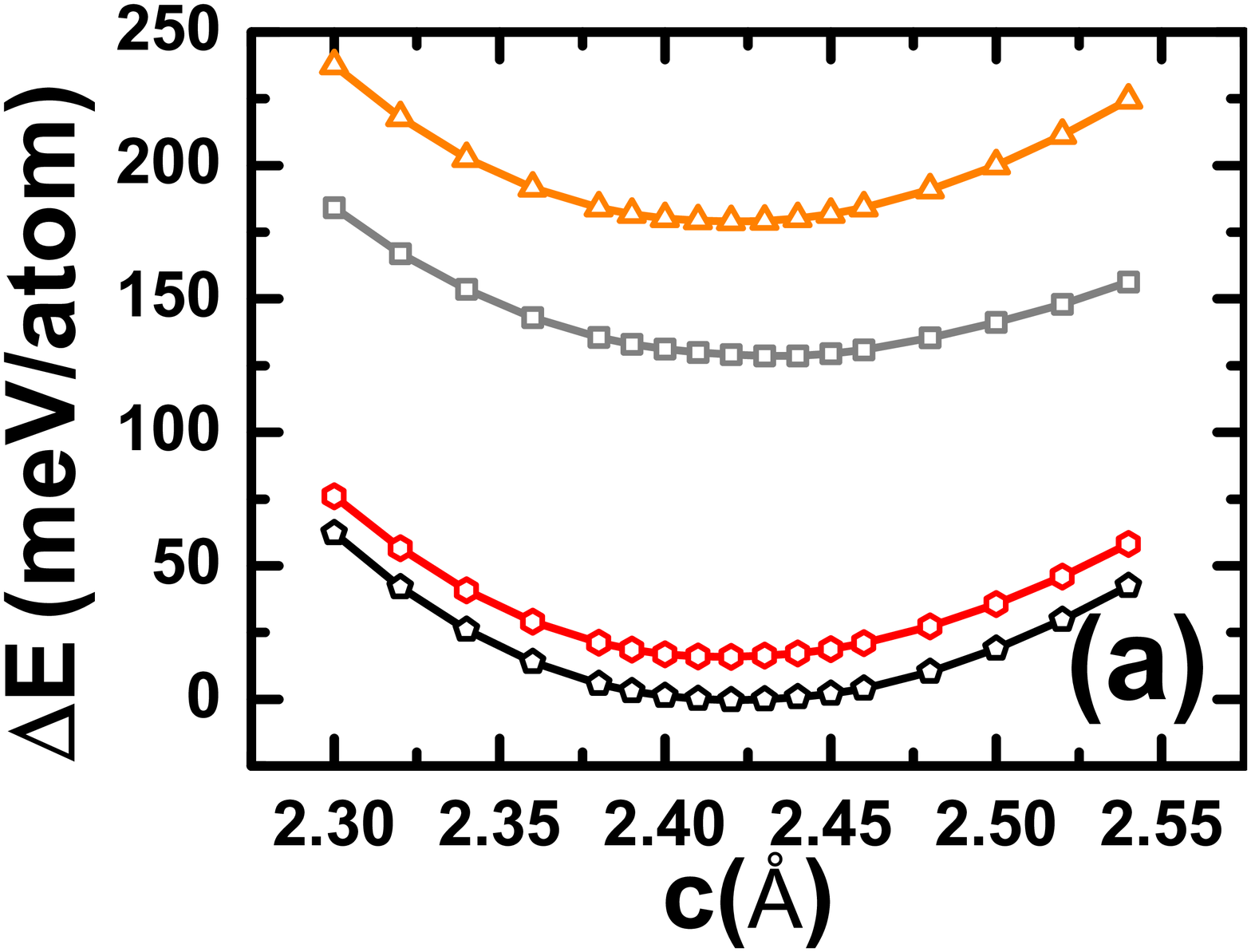}
\includegraphics[width=7cm]{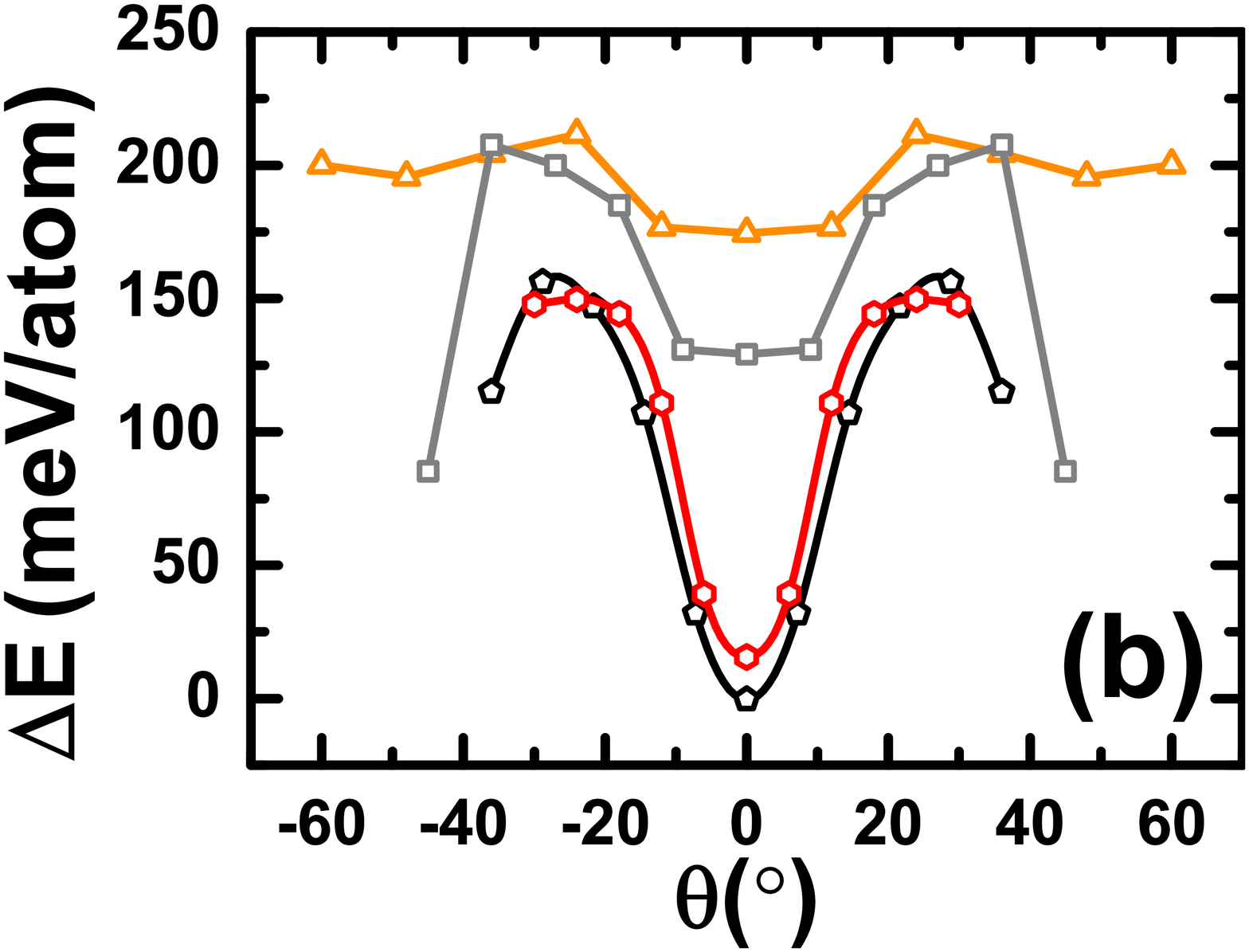}
\caption{{Relative total energy ($\rm \Delta E$) (in meV/atom) of
nanowires (in pristine form) as a function of (a)
the interplanar distance (with a fixed untwisted configuration) and
(b) the twisting angle. The figure presents results for
nanowires with triangular ($\triangle$), square ($\Box$),
pentagonal ($\pentagon$), and  hexagonal ($\hexagon$) cross sections.
Here, the reference energy is -4.732 eV/atom, corresponding to
an untwisted pentagonal nanowire.}}
\label{fig3}
\end{figure}

Figure \ref{fig3}(a) shows the nanowire relative total energy (per atom),
in their pristine forms, as a function of the lattice parameter $c$ (for untwisted
nanowires). The results indicated that nanowires present an elastic behavior for
up to about 10 \% of stretching deformation. The wires with pentagonal and hexagonal
cross sections have the lowest energies, being the most stable ones, with a total
energy of about 0.46 eV/atom higher than the value for silicon in a diamond cubic
phase (computed using the same theoretical framework). Those results suggest
that the nanowire configurations studied here are physically relevant, and have
energies that are comparable or lower than the the ones in configurations studied
elsewhere \cite{durgun,palaria}.

\begin{figure}[tb]
\includegraphics[width=8cm]{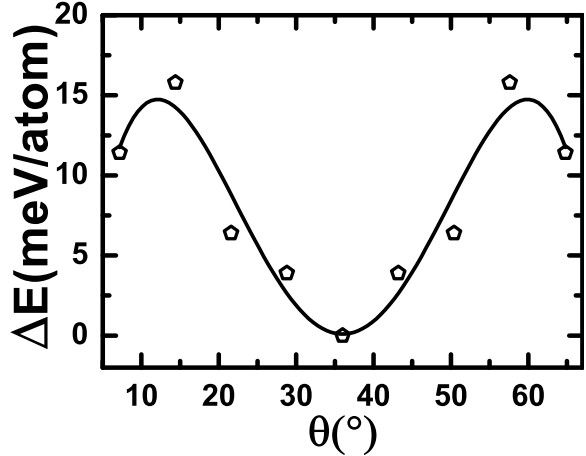}
\caption{{Relative total energy ($\rm \Delta E$) (in meV/atom) of
twisted pentagonal nanowire (in hydrogenated form) as a function of
the twisting angle.Here, the reference energy is -4.020 eV/atom.}}
\label{fig4}
\end{figure}

Figure \ref{fig3}(b) shows {the nanowire relative total energy (per atom)},
in their
pristine form, as a function of the twisting angle. The results indicated that,
for a wire of any cross section, the equilibrium state was associated to an
untwisted configuration. Additionally, the total energies of the pentagonal and
hexagonal nanowires presented a parabolic behavior (as a function of the
twisting angle), with an elastic limit which surpassed 10$^0$ of twisting.
Therefore, ultrathin nanowires follow a Hooke's law for torsion, even for
large angles. For hydrogenated nanowires, we observed a considerably smaller
twisting energy, as compared to the respective wires in their pristine forms.
Additionally, the energy minimum of the hydrogenated nanowires were associated
to twisted configurations. Therefore, hydrogenation drives a pristine nanowire on
a spontaneous transition toward a twisted configuration. This result is consistent
with what has been observed in several nanostructures, such as hydrogenation of
graphene nanoribbons \cite{koskinen}. Figure \ref{fig4} shows the {relative
total energy of
the hydrogenated pentagonal nanowire}, with an equilibrium configuration at an angle
of 36$^0$, and an elastic behavior for twisting angles around that minimum.

\begin{figure}[tb]
\includegraphics[width=8cm]{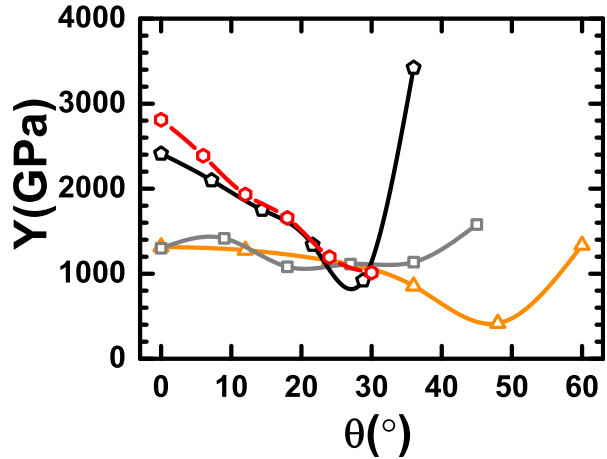}
\caption{Young's modulus of ultrathin silicon nanowires (in pristine form) as
a function of the twisting angle.
{The figure presents results for
nanowires with triangular ($\triangle$), square ($\Box$),
pentagonal ($\pentagon$), and  hexagonal ($\hexagon$) cross sections.}}
\label{fig5}
\end{figure}

The nanowire elasticity is described by the Young's modulus ($\rm Y$).
At a certain twisting angle, it is computed according to the respective curve
of total energy as function of the deformation strain, as given in
figure \ref{fig3}(a). This is obtained according to the expression
$\rm Y = (1/V_0 )(\partial ^2 E/\partial \varepsilon ^2 )_{\varepsilon  = 0}$,
where E is the strain energy, $\varepsilon$ is the axial strain, and $\rm V_{0}$
is the equilibrium volume, defined as the product of axial equilibrium length
($\rm L_{0}$) and the cross-sectional area ($\rm S_{0}$) of a nanowire.
For very thin nanowires, the definition of this area is very inaccurate,
which can affect the results. In order to compute the Young's modulus, each nanowire
was slightly elongated and contracted (up to a few percents) from its equilibrium
length $\rm L_{0}$, followed by full structural optimization \cite{lee}. The Young's
modulus was then determined by a third-order polynomial
expansion that fitted the calculated energy-strain curves. Table \ref{tab1}
presents the Young's modulus of all the nanowires studied here. Figure \ref{fig5}
shows the Young's modulus of nanowires, with all types of cross sections, as a
function of the twisting angles. There is a general trend that the Young's
modulus decreases with increasing of the twisting angle, although for the
pentagonal nanowire the Young's modulus increases considerably for twisting
angles beyond 30$^0$. This decrease in $\rm Y$, as result of twisting, has its origin
on the major changes on the chemical nature of inter-layer interatomic bonding.
The values of the Young's modulus could be compared to
available data for those of ultrathin nanowires \cite{xia}. The results for
ultrathin nanowires in their pristine form are larger than the ones
observed experimentally for several one dimensional nanostructures \cite{xia}.
For nanowires in their hydrogenated forms, as shown in figure \ref{fig1}(e),
the Young's modulus is considerably smaller than the respective ones in their
pristine forms, as shown in table \ref{tab1}. Those values are consistent with
results determined for hydrogenated nanowires \cite{zhu,kang,kou}.

The results indicated that nanowires with pentagonal and hexagonal cross sections
behave elastically for large twisting deformation, opening several possibilities
for applications as electromechanical devices in nanometrology.
It is particularly appealing to incorporate any type of silicon nanostructure,
as a probe nanodevice, in an integrated circuit within the current silicon device
technology \cite{garcia,waggoner}. An ultrathin nanowire could be used as a torsion
nanobalance or electrometer, similar to macroscopic torsion balances which have
been used to measure forces with great precision \cite{gund,cleland}. In fact, there
is growing interest in developing devices to measure forces with precision below the
10 $\rm \mu N$ limit \cite{naka,kramar2003,kim2010,gavartin}.

Working as a electrometer resonator with the two ends fixed in a holder, the
twisted nanowire would have a fundamental torsional resonance frequency
($\rm f_0$), which depends on the nanowire inertia moment ($I$) and the nanowire
angular spring constant ($G_o$). The angular spring constant, obtained from results
presented in  figure \ref{fig3}(b), is of the same order of the constant obtained
for ultrathin Mo$_6$S$_6$ nanowires (where deformation was presented in terms of the
twisting angle in deg/nm and potential energy in eV/nm). With the results from our
calculations, presented in figure \ref{fig3}(b) and table \ref{tab1}, a pentagonal
silicon nanowire in pristine form, with 20 layers and two fixed ends, would have
a resonance frequency of the order of about 80 GHz. This frequency is considerably
larger than the typical ones obtained in recent torsional experiments, in which
micrometer-size  paddles gave a resonance frequency of 2.61 MHz \cite{cleland},
but consistent with the size effects.

Following the idea of measuring variations in the oscillating frequencies due to
the presence of charged particles in the neighborhood of the twisting resonator
paddle device \cite{cleland}, a proper detection of frequency variations, for the
silicon nanowires here investigated, could be hampered by thermal noise, even at
very low temperatures. However, those ultrathin nanowires could have considerably
smaller resonance frequencies, which would favor those measurements. Reducing such
frequencies could be achieved by functionalizing the nanowires, such as by
hydrogenation or anchoring other functional groups, which would reduce considerably
the nanowire angular spring constant. Our results show that hydrogenation, as shown
in figure \ref{fig1}(e), leads to a major decrease in the $G_o$ by almost an order
of magnitude (according to results in figure \ref{fig4}), as compared to the same
nanowire in a pristine form. The physical origin of this behavior is the major change
on the chemical nature of silicon-silicon interactions as result of hydrogenation.

Our results suggest broad opportunities to obtain considerably high quality factors
and sensitivity to measure charged systems in the nanowire neighborhood \cite{cleland}.
A twisted nanowire could be used as a device to measure torsions at nanoscopic
level, with potential  applications to study several molecular properties of
DNA's \cite{bryant}. For a silicon pentagonal nanowire paddle with fixed ends,
described in the previous paragraph, torsion of a few degrees could allow to
explore torques and forces in the order of $\rm 10^{-20} {\rm N \, m}$ and 10 pN,
respectively.

\begin{figure}[tb]
\centerline{\includegraphics[width=100mm, angle=0.0]{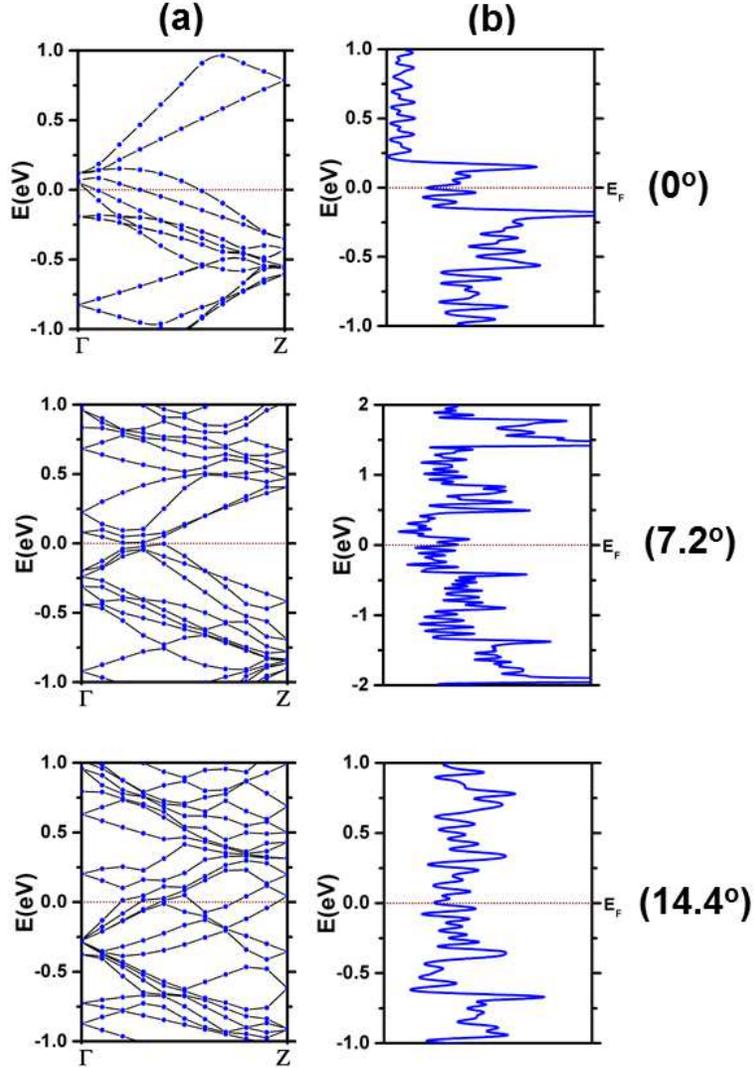}}
\caption{{Electronic band structure (a) and DOS (b) of pentagonal
silicon nanowires (in the pristine form) for some
twisting angles, near the Fermi level ($\rm E_F$).
Energies are given in eV and DOS in arbitrary unities.}}
\label{fig6}
\end{figure}

Our results indicate that the elastic properties of nanowires could be used to
probe forces and torques at molecular level. The angular spring
constant could be tailored to appropriate values for each application, by
anchoring functional molecules in the nanowires. However, the major challenge
would be how to translate the mechanical angular deformation of the nanowires in
terms of macroscopic measurements. One of the most appealing possibilities is to
explore the correlation between nanowire deformation and electronic transport,
a nanowire property which is generally easily accessible \cite{appenzeller}.
In order to study the changes on the electronic properties of the nanowires
as result of twisting, we computed the nanowire electronic band structures.
Figure \ref{fig6} presents the band structure of pentagonal nanowires, showing
that the nanowires are metallic for all twisting angles. Several bands cross the
Fermi level, which gives rise to a reasonably high density of states (DOS) in
that region. On the other hand, there is a large increase on the density of states
across the Fermi level as the torsion angle is increases, which would lead to
an increase in the wire conductivity with increasing the twisting angle.

\pagebreak

In summary, ultrathin twisted silicon nanowires, in pristine and hydrogenated forms,
have been investigated as potential candidates for mechanical torsional nanodevices.
All ultrathin nanowires followed a Hooke's law for stretching and twisting, and
such elastic properties could be explored to build nanoscopic torsion electrometers
to measure forces and torques at molecular level. Additionally, the torsion properties
could be tailored by functionalizing the nanowires.
{It should be stressed that nanowires of other materials, in pristine or
functionalized forms, are also potential candidates to work as torsional
electromechanical nanodevices.}

\acknowledgments
The authors acknowledge support from Brazilian Agencies FAPESP and CNPq. The
calculations were performed at the computational facilities of
LCCA of the University of S\~ao Paulo.


\end{document}